\documentclass[aps,prl,twocolumn,showpacs,floatfix,superscriptaddress]{revtex4}
\usepackage{dcolumn}
\usepackage{amsmath}
\usepackage{amsfonts}
\usepackage{amssymb}
\usepackage{epsfig}

\begin{document}

\title{Insights from ARPES for an undoped, four-layered, two-gap high-$T_{c}$
superconductor}
\author{Wenhui Xie}
\author{O. Jepsen}
\author{O. K. Andersen}
\affiliation{Max-Planck Institut f\"{u}r Festk\"{o}rperforschung, Heisenbergstr. 1,
D-70569 Stuttgart, Germany}
\author{Yulin Chen}
\author{Z.-X. Shen}
\affiliation{Stanford Synchrotron Radiation Laboratory, Stanford University, Stanford,
Ca, USA}
\date{\today }

\begin{abstract}
An undoped cuprate with apical fluorine and inner ($i$) and outer ($o$) CuO$%
_{2}$-layers is a 60\thinspace K superconductor whose Fermi surface (FS) has large $n$
and $p$-doped sheets with the SC gap on the $n$-sheet twice that on the $p$%
-sheet \cite{Chen}. The Fermi surface is \emph{not} reproduced by the LDA,
but the screening must be substantially reduced due to electronic
correlations, \emph{and }oxygen in the $o$-layers must be allowed to dimple
outwards. This charges the $i$-layers by 0.01$\left\vert e\right\vert $,
causes an 0.4 eV Madelung-potential difference between the $i$ and $o$%
-layers, quenches the $i$-$o$ hopping, and localizes the $n$-sheets onto the 
$i$-layers, thus protecting their $d$-wave pairs from being broken by
scattering on impurities in the BaF layers. The correlation-reduced
screening strengthens the coupling to $z$-axis phonons.
\end{abstract}

\pacs{71.18.+y,74.25.Jb,74.72.Jt}
\maketitle

Despite 20 years' intensive research, high-temperature superconductivity in
the cuprates has not been understood. Much insight has been gained through
discoveries of new cuprates and refinements of experimental techniques. An
example is the discovery \cite{Iyo} of an undoped 4-layered cuprate with
apical fluorine, Ba$_{2}$Ca$_{3}$Cu$_{4}$O$_{8}$F$_{2},$ which is not a Mott
insulator, but a 60\thinspace K HTSC; the experimental findings about
its Fermi surface and SC gap are reported in the preceding Letter \cite{Chen}.

All HTSCs consist of CuO$_{2}$-layers separated by apical blocks into which
dopants may be inserted. Undoped cuprates (Cu 3$d^{9})$ are antiferromagnetic
(AF) Mott insulators. Upon hole-doping beyond $p\mathrm{\sim }0.05$ per CuO$%
_{2}$, they become HTSCs with pairing symmetry $d_{x^{2}-y^{2}},$ and the
critical temperature reaches a maximum, $T_{c\,\max },$ when $p\mathrm{\sim }%
0.2$. For optimally and over-doped materials the Fermi surfaces (FS)
measured by ARPES enclose \emph{large} hole-areas of size $1+p$ (in units of
half the Brillouin-zone area)$,$ are centered at $\left( k_{x},k_{y}\right) 
\mathrm{=}\left( \pi ,\pi \right) ,$ and agree surprisingly well with
detailed predictions from LDA band-structure calculations \cite{ShenReview}.
Even the existence of a small sub-band splitting in bi-layered cuprates \cite%
{Sudhib,JPCS95}, which is twice the integral for inter-layer hopping, $%
t_{\perp }\left( \mathbf{k}_{\shortparallel }\right) \sim t_{\perp }\left(
\cos k_{x}-\cos k_{y}\right) ^{2}/4,$ and the existence of a $k_{z}$%
-dispersion proportional to $t_{\perp }\left( \mathbf{k}_{\shortparallel
}\right) \cos \frac{1}{2}k_{x}\cos \frac{1}{2}k_{y}\cos \frac{c}{2}k_{z}$ in
body-centered tetragonal materials \cite{Pavarini,Bansil}, have recently
been confirmed by ARPES \cite{Kordyuk,ShenReview} and angular
magnetoresistance oscillations (AMRO) \cite{Hussey}. Hence, over- and
optimally doped cuprates appear to have well-defined quasi-particle bands
with mass-renormalizations 2-4, not only in the direction of the layers, but
also in the perpendicular direction \cite{ShenReview,Kim,Lichtenstein}. In
the superconducting state, the magnitude of the gap is approximately 5.5$%
k_{B}T_{c}.$ For under-doped materials, a pseudogap opens up when $T<T^{\ast
}.$ The remaining Fermi arcs \cite{ShenReview,Argonne}, when mirrored around
the AF zone boundary, enclose \emph{small} hole-areas of size $p,$ are
centered at $\left( \frac{\pi }{2},\frac{\pi }{2}\right) ,$ and are roughly
reproduced by LDA+$U$ calculations with $U_{dd}\mathrm{\sim }6$\thinspace eV
and AF long-range order, or similar Hartree-Fock (HF) solutions for 3-band
models \cite{JPCS95,BansilRIXS}. Upon doping with electrons beyond a certain
level, a metallic AF (AFM) phase develops and, subsequently, a
superconducting phase with the electrons occupying \emph{small,} $\left( \pi
,0\right) $-centered FS areas of size $n.$ For higher $n$-doping \cite%
{Armitage,ShenReview}, antiferromagnetism disappears and the FS again
becomes $\left( \pi ,\pi \right) $-centered and large, of size $1-n$.
Optimal $n$-doping occurs for $n\mathrm{\sim }0.2$, and known values of $%
T_{c~\max }$ are much smaller $\left( \lesssim 40\,\text{K}\right) $ than
for optimal $p$-doping $\left( \lesssim 140\,\text{K}\right) $, and so are
the gaps. So far it has not been possible to $p$- and $n$-dope the same
compound.

\begin{figure}[tbp]
\epsfig{file=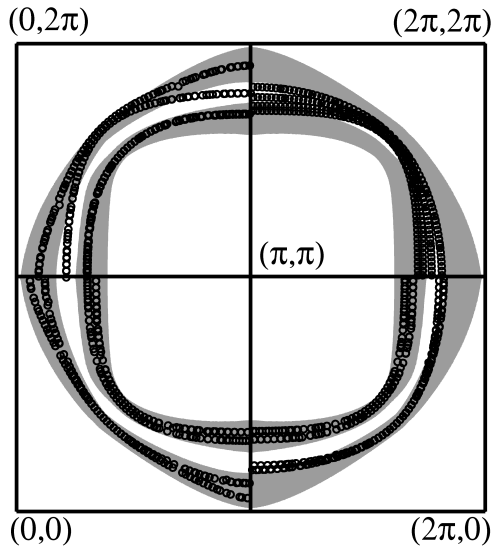,width=.9\linewidth,clip=true}
\caption{FS for Ba$_{2}$Ca$_{3}$Cu$_{4}$O$_{8}$F$_{2}.$ \textit{Grey:} ARPES 
\protect\cite{Chen}. The width reflects the experimental error bar. The
smaller (larger) sheet is the $n\left( p\right) $-sheet. \textit{Open dots: }%
Calculations.\textit{\ Top right:} LDA. \textit{Remaining three quarters:} $%
\langle $LDA+$U$$\rangle $. \textit{Bottom right:} Flat CuO$_{2}$-layers and
2\% F-O exchange ($q\mathrm{=}0.02)$ treated in the VCA. Indistinguishable
from this FS are the ones calculated with no F-O exchange, but with O in the 
$o$- or $i$-layers dimpled outwards by 0.07\thinspace \AA . See
Fig.\thinspace \protect\ref{FigPot}. The $n\left( p\right) $-sheet resides
on the $i\left( o\right) $-layers. $\protect\varepsilon _{F}$ is adjusted
such the FS fits along the nodal line whereby $n\mathrm{=}0.19$ and $p%
\mathrm{=}0.12.$ \textit{Bottom left:} Dimpling as before, but with the
vertical Ba and F positions adjusted such as to give the best agreement with
the ARPES FS: $z_{\mathrm{Ba}}-z_{\mathrm{Cu}_{o}}$ is decreased from 2.0 to 
$1.7$\AA\ and $z_{\mathrm{F}}-z_{\mathrm{Cu}_{o}}$ is decreased from 2.4 to $%
2.1$\AA . $\protect\varepsilon _{F}$ is adjusted as before, whereby $n%
\mathrm{=}0.18$ and $p\mathrm{=}0.21.$ \textit{Top left:} As before but with
O in the $o$- (or $i$-) layers dimpled \emph{inwards} by 0.10\thinspace \AA %
. Now, the $n\left( p\right) $-sheet resides on the $o\left( i\right) $%
-layers. $z_{\mathrm{Ba}}-z_{\mathrm{Cu}_{o}}=1.6$\AA\ and $z_{\mathrm{F}%
}-z_{\mathrm{Cu}_{o}}=2.0$\AA . $\protect\varepsilon _{F}$ is adjusted as
before, whereby $n\mathrm{=}0.15$ and $p\mathrm{=}0.18$. Note that
inter-sheet splittings in $\mathbf{k}$-space are enhanced near the
saddlepoints in the band structure.}
\label{FigFS}
\end{figure}

In Fig.\thinspace \ref{FigFS}, we show the FS of undoped Ba$_{2}$Ca$_{3}$Cu$%
_{4}$O$_{8}$F$_{2}$. ARPES resolves the two sheets shown in grey \cite{Chen}%
. Both are $\left( \pi ,\pi \right) $-centered and \emph{large,} their areas
being $1\pm p$ with $p\mathrm{\sim }0.2$. We shall refer to the larger one as
the hole-doped or $p$-sheet and to the smaller one as the electron-doped or $n$%
-sheet. Ba$_{2}$Ca$_{3}$Cu$_{4}$O$_{8}$F$_{2}$ thus appears to be the first 
\emph{self}-doped HTSC having $n$- and $p$-doping in one and the same
crystal. All other known undoped cuprates, single- as well as multi-layered,
are insulating, and when hole-doped, the observed FS-area splittings \cite%
{Kordyuk} are proportional to $t_{\perp }\left( \mathbf{k}_{\shortparallel
}\right) $ and almost an order of magnitude smaller than the $\pm 0.2$ found
in Ba$_{2}$Ca$_{3}$Cu$_{4}$O$_{8}$F$_{2}.$ The question we shall try to answer
in this letter is: How and why does this compound manage to dope itself?

There are no FS measurements for multi-layered cuprates with \emph{in}%
equivalent inner $\left( i\right) $ and outer $\left( o\right) $ layers,
with the exception of Ba$_{2}$Ca$_{3}$Cu$_{4}$O$_{8}$F$_{2},$ but \emph{%
site-selective} NMR studies \cite{NMR} indicate that $i$-layers have less
holes than $o$-layers and that superconductivity develops differently with
temperature in the different layers. This behaviour is interpreted in terms
of electrons localized onto individual layers, each of which follows the
generic phase diagram as a function of doping, and which are weakly coupled
by proximity effects \cite{MoriMaekawa}. Most spectacular are the properties
of 5-layered HgBa$_{2}$Ca$_{4}$Cu$_{5}$O$_{12+\delta }$ \cite{AFSC}. Here,
the $o$-layers form an optimally doped HTSC with $T_{c\,\max }\mathrm{=}108$%
\thinspace K and the three $i$-layers are AF \emph{metals} (AFM) with
moments of about 0.3$\mu _{B}.$ By decreasing the hole doping, $T_{c}$
decreases to 70\thinspace K and the $o$-layers form a \emph{uniformly} mixed
HTSC-AFM phase with $0.1\mu _{B}$ moments while the $i$-layers form a $%
0.7\mu _{B}$-AFM. The layers are flat and the estimated hole-counts for the $%
o$- and $i$-layers differ by as much as $0.2$ for the over-doped material.

Seen on the background of these truly exotic properties, the discovery \cite%
{Chen} that Ba$_{2}$Ca$_{3}$Cu$_{4}$O$_{8}$F$_{2}$ is the first HTSC with 
\emph{different SC gaps} on the two sheets, may seem less spectacular.
However, it is the gap on the $n$-sheet which is the largest, actually
twice the gap on the $p$-sheet, and this is opposite to what one would
expect from the generic phase diagram. Moreover, two-gap superconductivity
has only been observed unambiguously in a few materials of which MgB$_{2}$
is the most spectacular. There, the mechanism is conventional \cite%
{MgB2,Boeri}.

Whereas the FS sheets observed in all other multi-layered HTSCs are hardly
split along the nodal direction, Ba$_{2}$Ca$_{3}$Cu$_{4}$O$_{8}$F$_{2}$
exhibits a \emph{large nodal splitting, }$\Delta k_{\left[ 110\right] }.$
This is the \emph{third} anomaly of the ARPES data and the key to the
self-doping and the two gaps. For comparison with the ARPES FS, we show in
the top right-hand quarter of Fig.\thinspace \ref{FigFS} the LDA FS which
is seen to have four FS sheets, split almost exclusively by inter-layer
hopping \cite{LAPWdetails}. The four sub-bands are split by approximately $%
\pm \frac{1}{2}\left( \sqrt{5}\pm 1\right) t_{\perp }\left( \mathbf{k}%
_{\shortparallel }\right) $ with $t_{\perp }\mathrm{\sim }0.2\,\mathrm{eV}.$
This is the first case known to us where there is a substantial \emph{%
discrepancy} between an experimental, large FS and the LDA. The latter fails
to reproduce the $\Delta k_{\left[ 110\right] }\,\partial \varepsilon
/\partial k_{\left[ 110\right] }=0.4$\thinspace eV large \emph{%
Madelung-potential difference} (on electron scale) between the layers on
which respectively the $p$- and the $n$-nodal quasi-particles are located.
We have taken $\partial \varepsilon /\partial k_{\left[ 110\right] }$ as the
bare (LDA) nodal velocity, which is the consistent choice in the present
context \cite{footnote}. In the LDA (GGA), the electron potential averaged
over an $i$-layer is 0.024\thinspace eV (0.034\thinspace eV) less than that
averaged over an $o$-layer, that is 17 (12) times too small! The discrepancy
remains after we optimize the structure using the LDA. Thereby, most
notably, the oxygens in the $o$-layers dimple outwards by $\mathrm{\sim }3%
{{}^\circ}%
$, about one third the dimple in YBa$_{2}$Cu$_{3}$O$_{7}$ \cite{Dimpling}.

Since the 0.4\thinspace eV potential exceeds $\frac{1}{2}\left( \sqrt{5}%
+1\right) t_{\perp },$ it effectively \emph{blocks the hopping} between the $%
i$- and the $o$-layers. Therefore, the $n$-sheets are associated with the $i$%
-layers and the $p$-sheets with the $o$-layers, or the other way around. The
potential difference thus blocks inter-band scattering and protects the $i$%
-bands from scattering on the ubiquitous impurities in the apical
BaF-blocks. This may be the reason why two SC gaps survive and --in case the 
$p$-sheets reside on the $o$-layers-- also the reason why the gap on the $p$%
-sheets is the smaller one: it is suppressed by scattering on impurities in the
BaF-block.

Let us set this 0.4\thinspace eV potential difference in perspective: With
the in-plane lattice constant 3.86\thinspace \AA\ and the distance between
Ca and CuO$_{2}$ layers 1.54\thinspace \AA , Poisson's equation says that
the layer-averaged potential is a sawtooth with kinks of $\sim $20\thinspace
eV$\,Q_{n}$ on the layers. $Q_{n}$ is the charge per 2D cell. If Ba$_{2}$Ca$%
_{3}$Cu$_{4}$O$_{8}$F$_{2}$ were purely ionic, the potential would therefore
be 0 on the Ca$^{2+}$ layers and 20\thinspace eV on the $\left( \text{CuO}%
_{2}\right) ^{2-}$ layers, the \emph{same} on the $i$- and $o$-layers
because the CaCuO$_{2}$-\emph{unit is neutral.} Next, we transfer from the
BaF layers $q_{i}$ positive charges to each of the $i$-layers and $q_{o}$
positive charges to each of the $o$-layers. This sets up a difference in
Madelung potential between $o$- and $i$-layers of 40$q_{i}$\thinspace eV,
which depends only on $q_{i},$ not on $q_{o}.$The ARPES value of 0.4 eV
(neglecting exchange-correlation contributions) therefore tells us that the $%
i$-layers have a net charge of either $0.01\left\vert e\right\vert $, in
which case the $n$-sheets reside on the $i$-layers$,$ or of $-0.01\left\vert
e\right\vert $, in which case the $p$-sheets are on the $i$-layers. In the
LDA (GGA), the charging is 17 (12) times smaller.

Crystallographically it is possible that $q$ of the two
layer-oxygens per cell are exchanged with fluorine, so that the formula is Ba%
$_{2}$Ca$_{3}($CuO$_{2-q}$F$_{q})_{4}$O$_{4q}$F$_{2-4q}.$ In fact, LDA
calculations with $q\mathrm{=}0.25,$ employing not only the virtual-crystal
approximation (VCA), but also supercells, \emph{do} reproduce the ARPES FS.
However, the position of the apical F\thinspace 2$p_{z}$ level is
4\thinspace eV below that of apical O\thinspace 2$p_{z},$ and such
impurities in the CuO$_{2}$-layers are very efficient in breaking $%
d_{x^{2}-y^{2}}$-wave pairs: For an allowed $T_{c}$-suppression of 10\%, we
find that $q$ cannot exceed 0.001, which is even smaller than the charging
of 0.01 needed to create the observed crystal-field splitting in the absence
of screening \cite{Antbp}. This rules out F-O exchange as an explanation for
the anomalous ARPES FS.

QMC calculations for a $t_{1u}$-band Hubbard model for C$_{60}$ have
revealed that the RPA describes metallic sceening well for $U/W\mathrm{%
\lesssim }2,$ but that once $U/W\mathrm{\sim }3,$ electronic correlations
reduce the screening by an order of magnitude \cite{Gunnarsson}. This seems
to be the problem with the LDA when applied to Ba$_{2}$Ca$_{3}$Cu$_{4}$O$_{8}
$F$_{2},$ which is even nearly 2D. One might think of reducing all
inter-layer hoppings as was recently done using the Gutzwiller approximation 
\cite{Mori06}, but this would not directly change the metallic inter-layer
screening, $1+2NM\sim 25,$ where $2N\mathrm{\sim }0.6$\thinspace
electrons/(eV$\cdot $CuO$_{2}$) is the intra-layer density of states at the
Fermi level and $M\mathrm{\sim }40\,$eV$\cdot $CuO$_{2}$/electron is the
inter-layer Madelung constant. Instead, we perform selfconsistent LDA+$U$
calculations for AF $\left( \pi ,\pi \right) $-order and subsequently
calculate band-structures and FSs for the charge potential, neglecting the
spin potential. This procedure we denote $\langle $LDA+$U$$\rangle $. The
result for the pure compound $\left( q\mathrm{=}0\right) $ is very similar
to that obtained by the LDA and shown in the top right-hand quarter of
Fig.\thinspace \ref{FigFS}. However, with an F-O exchange corresponding to $q%
\mathrm{=}0.02,$ we \emph{do} obtain the proper splitting along the nodal
line. This is shown in the bottom right-hand quarter of Fig.\thinspace \ref%
{FigFS}. The $\langle $LDA+$U$$\rangle $ procedure thus reduces the
screening of this perturbation to about 2, but the pair-breaking argument
why F-O exchange is an unlikely explanation for the anomalous ARPES FS
remains valid. So from now on, we take $q\mathrm{=}0.$

\begin{figure}[tbp]
\epsfig{file=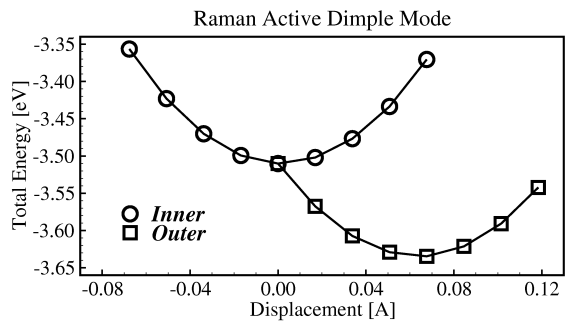,angle=0,width=.9\linewidth,clip=true}
\caption{Total energy as a function of the Raman-active dimpling
displacements calculated with LDA+$U.$ See Fig. \protect\ref{FigPot}.} 
\label{FigDimplingE}
\end{figure}

\begin{figure}[tbp]
\epsfig{file=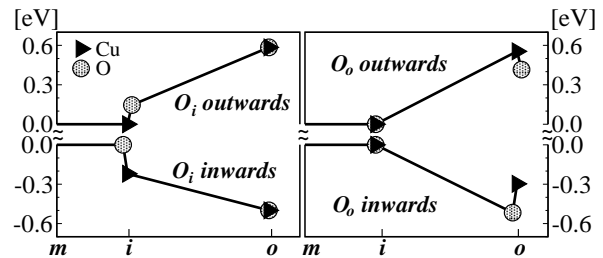,angle=-90,width=.9\linewidth,clip=true}
\caption{Layer-averaged potentials calculated with $\langle $LDA+$U$$\rangle 
$ for dimplings of $\pm $0.1\thinspace \AA\ of the inner (left) and outer
(right) layer-oxygens. This potential is the calculated Cu and O core
levels, lined up on the flat layer and normalized to zero at the Ca mirror
plane. $m$ is the mirror plane.}
\label{FigPot}
\end{figure}

Next, we include the \emph{calculated outwards dimple} of the \emph{outer} O$%
_{2}$-layer (see Fig.\thinspace \ref{FigDimplingE}), which is about the same
in the LDA and the LDA+$U.$ That this may induce a potential shift between
the $i$- and $o$-layers can be seen as follows: For the calculated
0.07\thinspace \AA $\,\sim \frac{1.54}{20}\,\mathrm{\mathring{A}}$ dimple,
the bare value of the Cu$^{2+}$-(O$^{2-})_{2}$ dipole in the $o$-layer is $%
\frac{3\times 20\,\mathrm{eV}}{20}\mathrm{=}3$\thinspace eV, and after
averaging over the Cu and O$_{2}$ sub-layers this yields a 2\thinspace eV
bare Madelung shift of the $o$-layer with respect to the $i$-layer.
Fig.\thinspace \ref{FigPot} shows the $\langle $LDA+$U$$\rangle $
layer-averaged potential for all the Raman-active (even) dimples of the $i$
and $o$-layers. We see that the intra-layer screening is substantial, that
the inter-layer screening is about 4, and that for the calculated
equilibrium structure, the $\langle $LDA+$U$$\rangle $ creates the proper
0.4\thinspace eV potential! The FS is very similar to the one calculated by
adjusting the F-O exchange to reproduce the nodal splitting of the ARPES FS
shown in the bottom right quarter of Fig.\thinspace \ref{FigFS}. Essentially
the same FS is obtained by dimpling the $i$-layers outwards. In all cases,
the $n$-sheets are on the $i$-layers and split by the hopping between them.
The $p$-sheets are on the $o$-layers and almost degenerate; they agree less
well with ARPES near the anti-nodal direction where the ARPES peaks are less
well defined. Moreover, the ARPES data has been forced to yield the area, $%
1\pm p,$ proper for zero doping, which for the real material may not hold
exactly. We have chosen to shift the Fermi level to fit ARPES along the
nodal direction. In the bottom left-hand quarter of Fig.\thinspace \ref%
{FigFS} we demonstrate that perfect agreement with ARPES can be obtained by
adjustment of the vertical Ba and F positions.

If --in disagreement with the total-energy result-- we dimple the $o$-layers
or the $i$-layers \emph{inwards} by 0.1\thinspace \AA ,\ the FS shown in
the top left-hand quarter of Fig.\thinspace \ref{FigFS} is obtained. Now the 
$n$-sheets are on the $o$-layers and unsplit, whereas the $p$-sheets are on
the $i$-layers and split by the hopping between them.

There has recently been interest in possible consequences of poor $z$-axis
screening for the electron-phonon interaction at small momentum transfers 
\cite{Olle,Devereaux}. We note that the coupling of the FS to an $a_{1g}$
Raman-active dimpling mode via the screened Madelung-potential difference
(once it exceeds the inter-layer hopping) is proportional to the Cu-O$_{2}$
distance, rather than to this distance squared, as is the case for the
hopping-induced coupling considered earlier \cite{CapeCod}. From
Fig.\thinspace \ref{FigPot} we realize that the deformation potential is
sizeable: $D\mathrm{=}5$\thinspace eV/\AA . For a rough estimate of the
associated electron-phonon coupling, let us like in Ref. \cite{Boeri} use
the Hopfield expression, $\lambda =ND^{2}/\left( M\omega ^{2}\right) ,$ to
compare with MgB$_{2}$ for which $\lambda \mathrm{\approx }1.0:$ In Ba$_{2}$%
Ca$_{3}$Cu$_{4}$O$_{8}$F$_{2}$, the density of states appropriate for $%
d_{x^{2}-y^{2}}$-pairing is twice the value in MgB$_{2},$ $D$ is 2.5 times
smaller, and $M\omega ^{2}$ is 1.3 times smaller for a Raman-active oxygen
dimpling mode than for the optical bond-streching mode in MgB$_{2}.$
Finally, the strong screening for large momentum transfers in Ba$_{2}$Ca$_{3}
$Cu$_{4}$O$_{8}$F$_{2}$ effectively limits the integrand to the central half
of the 2D Brillouin zone. As a result, $\lambda \mathrm{\sim }\frac{1}{3},$
which suffices to enhance some other mechanism for HTSC.

We have shown \emph{how} undoped Ba$_{2}$Ca$_{3}$Cu$_{4}$O$_{8}$F$_{2}$
manages to self-dope: By a dimpling distortion which is screened out poorly
due to strong electronic correlations. This sets up an 0.4\thinspace eV
difference in Madelung potential between the outer and inner layers, and
that causes the FS to split along the nodal direction. The large
Madelung-potential difference prevents interband scattering, and that seems
to be the reason why two SC gaps survive. Since the dimpling is outwards
according to our total-energy LAPW calculations, both in the LDA and in the
LDA+$U,$ the FS sheet with the small gap, the $p$-sheet, is located on the
outer layers, next to the BaF-blocks where impurities are ubiquitous. This
suggests that the $p$-gap is small because it is suppressed by impurity
scattering, while the $n$-sheet is protected by the Madelung potential. The
phase-diagram model, on the other hand, predicts the $p$-gap to be the
largest. A further reason for the different gap sizes could be that, like in
MgB$_{2}$ \cite{MgB2,Boeri}, (one of) the interaction(s) driving the
superconductivity is stronger for the $n$- than for the $p$-sheet. It could
be of relevance that the bonding $n$-sheet has a larger effective
intra-layer second-nearest neighbor Cu-Cu hopping integral, $t^{\prime }+%
\frac{1}{8}t_{\perp }\sim 0.39t,$ than the $p$-sheet, for which the
effective value is merely $t^{\prime }\sim 0.33t,$ and therefore nests
better along the $\left[ 10\right] $-directions. This would be consistent
with the observation that for hole-doped multi-layer cuprates $T_{c\,\max }$
correlates positively with the largest sheet-value of $t^{\prime }/t$ \cite%
{Pavarini}.

\emph{Why} these anomalies have been observed only in Ba$_{2}$Ca$_{3}$Cu$_{4}
$O$_{8}$F$_{2}$ is undoubtedly due to the apical ions being F rather than O.
Presumably, when going from apical O to the more ionic F, the covalency
between Ba and the apical ion is reduced, whereby the covalency between Ba
and O in the $o$-layer is strenghtened and causes it to dimple towards Ba.
This is supported by the experimental finding that the outer layers in HgBa$_{2}$Ca$_{4}$Cu$_{5}$O$%
_{12+\delta }$ are dimpled inwards \cite{Akimoto1997} or maybe flat \cite{AFSC,Huang1994}.

It will be interesting to follow the development of the Ba$_{2}$Ca$_{3}$Cu$%
_{4}$O$_{8}($F$_{1-2p}$O$_{2p})_{2}$ FS and its SC gaps as a function of
hole-doping $p.$ If our assignment of the $p$-sheets with the $o$-layers is
correct, and as long as screening remains poor, O doping should suppress the
gap on the $p$-sheet. Currently it is known that $T_{c\,\max }\mathrm{=}105$%
\thinspace K is reached at $p\mathrm{=}0.20$ \cite{Iyo}, but no ARPES data
are available.

We are grateful to O. Gunnarsson and J. Zaanen for useful discussions, to A.
Yamasaki for checking calculations.


\begin{thebibliography}{99}
\bibitem{footnote} Going $\Delta k_{\left[ 110\right] }$ away from a FS
sheet, the corresponding ARPES band first slopes $\sim $3 times less and
then, after a kink at $\sim $0.07\thinspace eV, about the same as the LDA
band.
\end{thebibliography}
\end{document}